\documentclass[pra,aps,showpacs,twocolumn]{revtex4-1}

\usepackage{amsfonts}
\usepackage{amsmath}
\usepackage{amssymb}
\usepackage{graphicx}

\newcommand{\be}{\begin{equation}}
\newcommand{\ee}{\end{equation}}
\newcommand{\ba}{\begin{eqnarray}}
\newcommand{\ea}{\end{eqnarray}}

\newcommand{\ketbra}[2]{\ensuremath{\vert{#1}\rangle\langle
{#2}\vert}}
\newcommand{\wh}{\widehat}
\newcommand{\refc}[1]{(\ref{#1})}

\begin{document}

\title{Unitary expansion of the time evolution operator}
\date{\today}
\author{ N. Zagury,$^1$ A. Arag\~ao,$^1$  J. Casanova,$^2$  E. Solano$^{2,3}$ } \affiliation{$^1$ 
Instituto de F\'{\i}sica, Universidade Federal do Rio de Janeiro. Caixa Postal 68528, 21941-972 Rio de Janeiro, RJ, Brazil \\
$^2$ Departamento de Qu\'{\i}mica F\'{\i}sica, Universidad del Pa\'{\i}s Vasco-Euskal Herriko Unibertsitatea, Apdo. 644, 48080 Bilbao, Spain \\
$3$ IKERBASQUE, Basque Foundation for Science, Alameda Urquijo 36, 48011 Bilbao, Spain}

\begin{abstract}
We propose an expansion of the unitary evolution operator, associated to a given Schr\"odinger equation, in terms of a finite product of explicit unitary operators. In this manner, this unitary expansion can be truncated at the desired level of approximation, as shown in the given examples.
\end{abstract}


\maketitle

\section{Introduction}

The central problem in any dynamical theory is to find the time evolution of a system that was prepared in a given initial state. In quantum mechanics there are only a few of these problems that are  readily solved by simple analytical methods~\cite{Cohen91}. In general, we  have to rely on approximations  to obtain out of the Schr\"odinger equation the time evolution operator $\widehat U_S(t,t_0)$ in a suitable form. With the explicit knowledge of $\widehat U_S(t,t_0)$, we may calculate the expectation value of any physical observable of the system at any time $t$ once we know the state of system at the time $t_0.$ Frequently, we are able to find the time evolution operator $\widehat U_0(t,t_0)$ associated with $\widehat H_0$, a part of the total Hamiltonian $\widehat H=\widehat H_0+\widehat H_{\rm int}. $  In this case, it is usually convenient to make a transformation to the ``interaction picture'' such that
\be \label{trans1}
\widehat U_S(t,t_0)=\widehat U_0(t,t_0)\widehat U_I(t,t_0)
\ee
holds with $\widehat U_0(t_0,t_0) = \widehat U_I(t_0,t_0)= \widehat{I}.$ 
Consequently, the  time evolution operator in the interaction picture, $ \widehat U_I(t,t_0), $ satisfies the Schr\"odinger equation  
\begin{equation}\label{Schro}
\frac{\partial \widehat{U}_I(t,t_0)}{\partial t}=-i \lambda \widehat{H}_1(t)\widehat{U}_I(t,t_0)\, ,
\end{equation}
where we have considered $\hbar=1$ and  have defined
$\lambda \widehat{H}_1(t)=\widehat U_0^\dagger(t,t_0) \widehat H_{\rm int} (t)\widehat U_0(t,t_0).$ The parameter $ \lambda$ is some dimensionless real parameter chosen to  give a measure of  the relative magnitude of the most important matrix elements of  $H_{\rm int}$ and $H_0$ in a given problem. 

The most popular perturbation approximation method to deal with the Schr\"odinger  equation in Eq.~(\ref{Schro}) is the Dyson expansion~\cite{Dyson49}:
\begin{eqnarray}\label{Sch}
&&\widehat{U}_I(t,t_0)= \widehat{I}-i \lambda \int_{0}^t dt_1\widehat{H}_1(t_1,t_0)+\nonumber\\
&&{(-i \lambda)^2} \int_{0}^t dt_1 \int_{0}^{t_1} dt_2
\widehat{H}_1(t_1,t_0)\widehat{H}_1(t_2,t_0)+... \, ,
\end{eqnarray}
where it is very convenient to estimate the solution by truncating the series to a given order $\lambda^n$. Besides the normal difficulties to calculate high-order terms, the Dyson truncation produces an approximated evolution operator that is not unitary. Other expansions of operator $\widehat U_1(t,t_0)$ have also been proposed in the literature, as the Magnus expansion~\cite{Magnus54} the Fer product~\cite{Fer58} and, more recently, the Aniello expansion~\cite{Aniello05}. New bounds for the convergence of the Magnus expansion and the Fer product have been recently studied in Ref.~\cite{Blanes98}. Other product expansions have also been considered in the literature~\cite{see}.

In this paper, we present an alternative  method to calculate the time evolution operator  $U_I(t,t_0)$ as a product of a finite number of unitary operators
\begin{equation}\label{expansion}
\widehat U_I(t,t_0)=\widehat U_1(t,t_0) \widehat U_2(t,t_0) \cdots \widehat U_k(t,t_0) \cdots \widehat U_N(t,t_0)\, ,
\end{equation}
where each operator $\widehat U_k(t,t_0), k=1,2,...N-1$, can be written as
the exponential of an anti-Hermitian operator proportional to $(\lambda)^k $, while $\widehat U_N(t,t_0) - \widehat{I}$ is at most of order $( \lambda)^N$. The number $N$ of operators  in the expansion can be as large as we want. If we approximate the last operator $\widehat U_N(t,t_0)$
in the product by the unit operator, we obtain an expansion of the evolution operator $\widehat U_I(t,t_0)$ which is explicitly unitary to order  $( \lambda)^{N-1}$.  Besides this important advantage, this expansion is well suited to treat a variety of problems. In Section II, we derive the expressions for each term in the expansion; in Section III, we provide pedagogical examples; and in Section IV, we present our conclusions.

\section{The method}

We start with the simple case of $N=2.$ Equation~\refc{expansion} can then be written as 
\be \label{trans2}
\widehat U_I (t,t_0)=\widehat U_1 (t,t_0)\widehat U_2 (t,t_0).
\ee
Following the same kind of procedure used in the transformation to the interaction representation, we write
\begin{equation}\label{W1}
\widehat{U}_{I}(t,t_0)=e^{-i \lambda\widehat{W}_{1}(t,t_0)} \widehat{U}_{2}(t,t_0)\, ,
\end{equation}
where   $\widehat{W}_{1}( t,t_0)$  is an hermitian operator to be chosen conveniently. From now on, we set $t_0=0 $ and avoid writing it to simplify the notation. From Eqs.~\refc{Schro}, \refc{trans2} and \refc{W1}, we have 
\be \label{Sch2}
\frac{\partial \widehat{U}_{2}(t)}{\partial t}=-i \lambda\widehat{H}_{2}(t)\widehat{U}_{2}(t)\, ,
\ee
where 
\begin{eqnarray}\label{H2}
\widehat{H}_{2}(t) \! = \! \sum_{m=0}^{\infty}\frac{(i \lambda)^{m}}{m!}(ad\hspace{0.05cm}\widehat{W}_{1}(t))^{m} \Big\{ \widehat {H}_{1}(t)
-\frac{1}{m+1}\frac{\partial \widehat{W}_{1}(t)}{\partial t} \Big\} . \nonumber \\ 
\end{eqnarray}
Here, we have defined $ ad \widehat A\{\cdot\}=[\widehat{A},\cdot]$ and used the following relation
\begin{eqnarray}
e^{\alpha \widehat{A}}\widehat{B}e^{-\alpha \widehat{A}} &=&\sum_{m=0}^\infty\frac{\alpha^{m}}{m!}(ad\hspace{0.05cm}\widehat{A})^{m} \big\{ \widehat{B} \big\} \nonumber\\
 &=&\widehat{B}+\alpha[\widehat{A},\widehat{B}]+\frac{\alpha^{2}}{2!}[\widehat{A},[\widehat{A},\widehat{B}]]+\cdots
\end{eqnarray}
Choosing  
\begin{equation}\label{w1}
 \widehat{W}_{1}(t)=\int_{0}^{t}\widehat{H}_{1}(t^\prime)dt^\prime \, ,
\end{equation}
we have 
\begin{equation}
\widehat{H}_{2}(t)=\sum_{m=1}^{\infty }\frac{(i \lambda) ^{m}}{(m+1)!}(ad\hspace{0.05cm
}\widehat{W}_{1}(t))^{m} \big\{m \widehat {H}_1(t) \big\} \, ,
\end{equation}
which is of order $ \lambda $. In this case, $\widehat{U}_2(t)$ is the solution of Eq.~\refc{Sch2} and should be written as an exponential of a non-Hermitian operator that is, in general, a series on the variable  $\lambda$, starting with $\lambda^2.$  

In the simple case where $[\widehat {H}_1(t),\widetilde {H}_{1}(t^\prime)]=-2if(t,t^\prime)$  is a c-number function, then
\be
\widehat H_2(t)=\lambda \int_0^t dt^{\prime} f(t^{\prime},t) 
\ee
is also a c-number function. Consequently, Eq.~\refc{Sch2} can be easily integrated to give $\widehat{U}_2(t)=e^{-i\lambda^2\phi(t)}$, where 
\be
\phi(t)= \int_0^t dt^\prime\int_0^{t^\prime}dt^{\prime\prime}f(t^{\prime\prime},t^\prime)
\ee
 and the time evolution operator is 
\begin{equation}\label{U1a}
\widehat U_I(t)= e^{ -i\lambda \int_{0}^{t} dt_1\widehat{H}_1(t_1)} e^{-i\lambda^2\phi(t)} .
\end{equation}
This result is  well known and could also be easily obtained  using  the Magnus expansion~\cite{Pechugas66}. It can be used, for example, to easily obtain the time evolution operator for the quantum state generated by an external time-dependent  force  acting on a mechanical oscillator, as we show in the next section. 

The procedure described above can be generalized for any value of $N$ greater than $2$ by setting 
\begin{equation}\label{1}
\widehat{U}_{n}(t)=e^{-i \lambda\widehat{W}_{n}(t)}\widehat{U}_{n+1}(t)\, , n=2,3...N-1\, ,
\end{equation}
so that expansion of the operator $\widehat U_I(t)$ may also read
\begin{equation}\label{expansion2}
\widehat U_I(t)= e^{-i \lambda\widehat{W}_{1}(t)}e^{-i \lambda\widehat{W}_{2}(t)}...e^{-i \lambda\widehat{W}_{N-1}(t)}\widehat U_N(t).
\end{equation}
 The operators $\widehat{U}_{n}(t),$ for $ n=2,3..$ satisfy a Schr\"odinger-like equation
\begin{eqnarray}\label{Sch3}
\frac{\partial \widehat{U}_{n}(t)}{\partial t}=-i \lambda\widehat{H}_{n}(t)\widehat{U}_{n}(t)\, ,
\end{eqnarray}
where $\widehat{H}_{n}(t)$ is given by 
\begin{eqnarray}\label{Hn}
\widehat{H}_{n}(t)&=&\sum_{m=0}^{\infty} \frac{(i \lambda)^{m}}{m!}(ad \hspace{0.05cm} \widehat {W}_{n-1}(t))^{m} \nonumber \\
&&\times\Big\{ \widehat{H}_{n-1}(t)
-\frac{1}{m+1}\frac{\partial \widehat{W}_{n-1}(t)}{\partial t} \Big\} \, .
\end{eqnarray}

By choosing  operators $\widehat{W}_{j}(t)$'s 
for $j=1,...N-1,$
we  obtain operators $\widehat{H}_j(t)$ for $j=2,...N$,  and the expansion given by Eq.~\refc{expansion2}. 

 We  now show that it is possible to choose operators $\widehat{W}_{n}(t)$ being  proportional to $\lambda^{n-1}$, and such that  the operators  $\widehat{H}_n(t)$ are  power series in the variable $\lambda$ starting with the power  $\lambda^{n-1}. $ Then, by substituting $\widehat{W}_{n}(t), n=1,2...(N-1)$ in Eq.~\refc{expansion} and noticing that  $\widehat{I} - \widehat{U}_N(t)$ would be  at least of ${\cal{O}} (\lambda^N),$   we will obtain the desired expansion announced in the Introduction. We make the proof by construction. Writing  explicitly the dependence of the operators  $\widehat{W}_{n}(t)$ and $\widehat{H}_{k}(t)$ on $\alpha=i  \lambda$, we have
\begin{eqnarray}\label{hypothesis}
\widehat{W}_{k}(t) &=&\alpha ^{k-1}\widetilde{W}_{k}(t) \,  ,\nonumber
\\
\widehat{H}_{k}(t) &=&\sum_{j=k-1}^{\infty }\widetilde{H}_{k,j}(t)\alpha ^{j}.
\end{eqnarray}
By substituting Eq.~\refc{hypothesis} in Eq.~\refc{Hn} we get,
for  $j\ge n\ge 1,$
\begin{widetext}
\begin{equation}\label{hn}
\widetilde{H}_{n+1,j}=\sum_{m=1}^{\infty }\sum_{k=1}^{n}\sum_{i=k-1}^{\infty }
\frac{1}{m!}(ad\widetilde{W}_{k}(t\hspace{0.05cm}))^{m} \Big\{ \widetilde{H}
_{k,i}\delta (km+i-j) \Big\} - \sum_{k=1}^{n}\sum_{m=0}^{\infty }\frac{1}{(m+1)!}(ad
\widetilde{W}_{k}(t) \hspace{0.05cm})^{m} \Big\{  \frac{d\widetilde{W}_{k}(t)}{d t} \delta (km+k-1-j) \Big\}  \, ,
\end{equation}
and for  $0<j<n,$
\begin{equation}\label{doublev}
\frac{d\widetilde{W}_{j+1}(t)}{dt}=\sum_{m=1}^{\infty
}\sum_{k=1}^{n}\sum_{i=k-1}^{\infty }\frac{1}{m!}(ad\widetilde{W}_{k}(t)
\hspace{0.05cm})^{m} \Big\{  \widetilde{H}_{k,i}\delta (km+i-j) \Big\}
- \sum_{k=1}^{n} \sum_{m=1}^{\infty }
\frac{1}{(m+1)!}(ad\widetilde{W}_{k}(t)\hspace{0.05cm})^{m} \Big\{ \frac{d\widetilde{
W}_{k}(t)}{dt}\delta (km+k-1-j) \Big\} \, ,
\end{equation}
\end{widetext}
where $\delta (m-n)= \delta_{m,n}$ is the Kronecker delta. Notice that $j+1{>}k$ in Eq.~(\ref{doublev}) so that $\frac{d
\widetilde{W}_{j+1}(t)}{dt}$ is given recursively in terms of the operators $\widetilde{W}_{k}(t)$ and $\widetilde{H}_{k}(t),$ for  $k\leq j$.
For example,
if we set $n=2$ and $j=1$  in the above equations, we easily get
\begin{eqnarray}\label{W2}
\frac{d\widetilde{W}_{2}(t)}{dt} &=&\frac{1}{2} ( ad\widetilde{W}_{1}(t)
\hspace{0.05cm}) \big\{ \widehat {H}_1(t) \big\}  \, ,
\end{eqnarray}
where we used Eq.~\refc{doublev} and the fact that $\widetilde{H}_{1,0}(t)=\widehat{H}_1(t).$ Using the initial condition
$\widehat{W}_{2}(0)=0, $ we have
\begin{eqnarray}\label{w2}
\widetilde{W}_{2}(t)&=&\frac{1}{2}\int_{0}^{t}dt^{\prime} [\widetilde{W}_{1}(t^{\prime} \hspace{0.05cm}),\widehat {H}_1(t^{\prime})]\, ,
\end{eqnarray}
which also can be written as
\begin{equation}
\widetilde{W}_{2}(t)=\frac{1}{2}\int_{0}^{t}dt_1\int_{0}^{t_1}dt_2 [\widehat{H}_1 (t_2),\widehat{H}_1 (t_1)]\, .
\end{equation}

To obtain an approximate expression for $\widehat U_I(t)$ valid to order {\cal O}$(\lambda^2),$
 we first set $N=3 $ in  Eq~\refc{expansion2}:
  \begin{equation}
 \widehat U_I(t) =e^{ -i\lambda\widetilde{W}_{1}(t)}e^{-(i\lambda)^2\widetilde{W}_{2}(t)} \widehat U_3(t)\, .
 \end{equation}
$\widehat U_3(t) - \widehat{I}$ is of order $\lambda^3,$  since it satisfies the Schr\"odinger equation, Eq.~\refc{Sch3}, with $\widehat H_3(t)$ of the order {\cal O}$(\lambda^2).$  If we approximate $\widehat U_3(t)$ by the identity  we get an approximation which is unitary and valid to order  ${\cal O}(\lambda^2).$ Using the expressions for
$\widetilde{W}_{1}(t)=\widehat{W}_{1}(t)$ and for $\widetilde{W}_{2}(t)$ given in Eq.~\refc{w1} and Eq.~\refc{w2}, we have
\begin{eqnarray} \label{U1}
&& \widehat U_I(t) \approx \exp\{ (-i\lambda) \int_{0}^{t}dt_1\widetilde H_1(t_1)  \}\times  \nonumber \\
&& \exp\{ \frac{\lambda^2}{2}\int_{0}^{t}dt_1\int_{0}^{t_1}dt_2 [\widehat{H}_1 (t_2),\widehat{H}_1 (t_1)]\}\,  .
\end{eqnarray}

The procedure described above  can be generalized for obtaining approximations involving a product of $N$ operators,
by calculating $\widetilde{W}_{k}(t), k=1,...N, $ through Eqs. \refc{hn} and \refc{doublev}. 
Below we give, as examples,  the explicit expressions for $\widetilde{W}_{3}(t),$ $\widetilde{W}_{4}(t),$
 and $\widetilde{W}_{5}(t)$
 \begin{eqnarray}
\widetilde{W}_{3}(t)&=&\frac{1}{3}\int_{0}^{t}dt^{\prime}[\widetilde{W}_{1}(t^{\prime}),
[\widetilde{W}_{1}(t^{\prime}),\widetilde {H}_{1}(t^{\prime})]]\, , \nonumber \\
\widetilde{W}_{4}(t) &=&\frac{3}{4!}\int_{0}^{t}dt^{\prime}[\widetilde{W}_{1}(t^{\prime}),[\widetilde{W}_{1}(t^{\prime}),[\widetilde{W}_{1}(t^{\prime}),
\widetilde {H}_{1}(t^{\prime})]]] \nonumber \\
&+&\frac{1}{4}\int_{0}^{t}dt^{\prime} [\widetilde{W}_{2}(t^{\prime}),[
\widetilde{W}_{1}(t^{\prime}),\widetilde {H}_{1}(t^{\prime})]] \, ,\nonumber \\
\widetilde{W}_{5}(t)&=&\frac{4}{5!}\int_{0}^{t}dt^{\prime}[\widetilde{W}_{1}(t^{\prime}), \nonumber \\ &&
[\widetilde{W}_{1}(t^{\prime}),[\widetilde{W}_{1}(t^{\prime}),[\widetilde{W}_{1}(t^{\prime}),
\widetilde {H}_{1}(t^{\prime})]]]]\nonumber \\
&+&\frac{2}{3!}[\widetilde{W}_{2}(t^{\prime}),[\widetilde{W}_{1}(t^{\prime}),
[\widetilde{W}_{1}(t^{\prime}),\widetilde {H}_{1}(t^{\prime})]]] .
\end{eqnarray}

As we show in the next section, the expansion obtained above may be useful in several cases and in particular for obtaining effective time independent hamiltonians, when the  operator 
$\widehat{U}_{n}(t^{\prime})$  in the expansion can be approximated by the exponential of the product of the time with a constant operator. 

Notice that  besides the fact that they are Hermitian,  no restriction was made on the operators $\widehat{W}_{n}(t)$ for $n=2,3...$ until now. Special choices of  $\widehat{W}_{n}(t),$ other than the one we have chosen to discuss in this paper,  may lead to interesting applications in specific cases.

\section{Examples of applications}

In this section we describe three  examples of applications of the method: i) the  problem  of a linear harmonic oscillator subjected to a driving force; ii) the Raman resonant transition inside a cavity; iii)  the ultrastrong coupling (USC) and deep strong coupling (DSC) regimes of the Jaynes-Cummings (JC) model. 

We start with the well known problem  of a linear harmonic oscillator subject to a driving force
$-g f(t).$ The Hamiltonian is  given by $(\hbar=1)$
\begin{equation}
H=\omega (\widehat a^\dagger \widehat a +1/2) + g f(t)(\widehat a +\widehat a^\dagger)
\end{equation}
where $\widehat a$ and $\widehat a^\dagger$ are the usual annhilation and creation operators satisfying the algebra $[\widehat a,\widehat a^\dagger]=1.$

 We first take $H_0=\omega (\widehat a^\dagger \widehat a +1/2)$ and go to the interaction representation by defining $\widehat U(t)=e^{-i\widehat H_0 t/\hbar}\widehat U_1(t)$, where
 \begin{equation}
\frac{\partial \widehat{U}_1(t)}{\partial t}=-i g \widehat{H}_1(t)\widehat{U}_I(t)\, ,
\end{equation}
 with
 \be
\widehat{H}_1(t)=f(t)(\widehat a e^{-i\omega t} +\widehat a^\dagger e^{i\omega t}) .
 \ee
In this case, $[\widehat{H}_1(t),\widehat{H}_1(t^\prime )]=-2if(t)f(t^\prime)\sin{\omega(t-t^\prime) }$ is a c-number. Therefore,  $\widehat{W}_n = 0$ for $ n>2,$  $\widehat{U}_3 = \widehat{I},$  and
\ba
\widetilde{W}_1(t)&=& \int_{0}^{t}dt^\prime f(t^\prime)(\widehat{a} e^{-i\omega t^\prime} + \widehat{a} ^\dagger e^{i\omega t^\prime}) ,  \\
\widetilde{W_2}(t)&=& i\int_{0}^{t}dt_1\int_{0}^{t_1}dt_2 f(t_1)f(t_2)\sin{\omega(t_1-t_2 )}.\nonumber
\ea
Then, the  time evolution operator in the  interaction picture is given by
\be
\widehat U_1(t)= e^{i\varphi(t)} \widehat{D}(v(t))
\ee
where $\varphi(t)=g^2\widetilde{W}_2(t)$ is a time-dependent phase and $ \widehat{D}(v(t))=e^{v(t)\widehat a^\dagger -v^*(t)\widehat a}$  is the displacement operator and
 \be
v(t)=-i g \int_{0}^{t}dt^\prime f(t^\prime)e^{i\omega t^\prime}.
\ee

\begin{figure}[ht]
 \label{Raman}
\includegraphics[width=6cm]{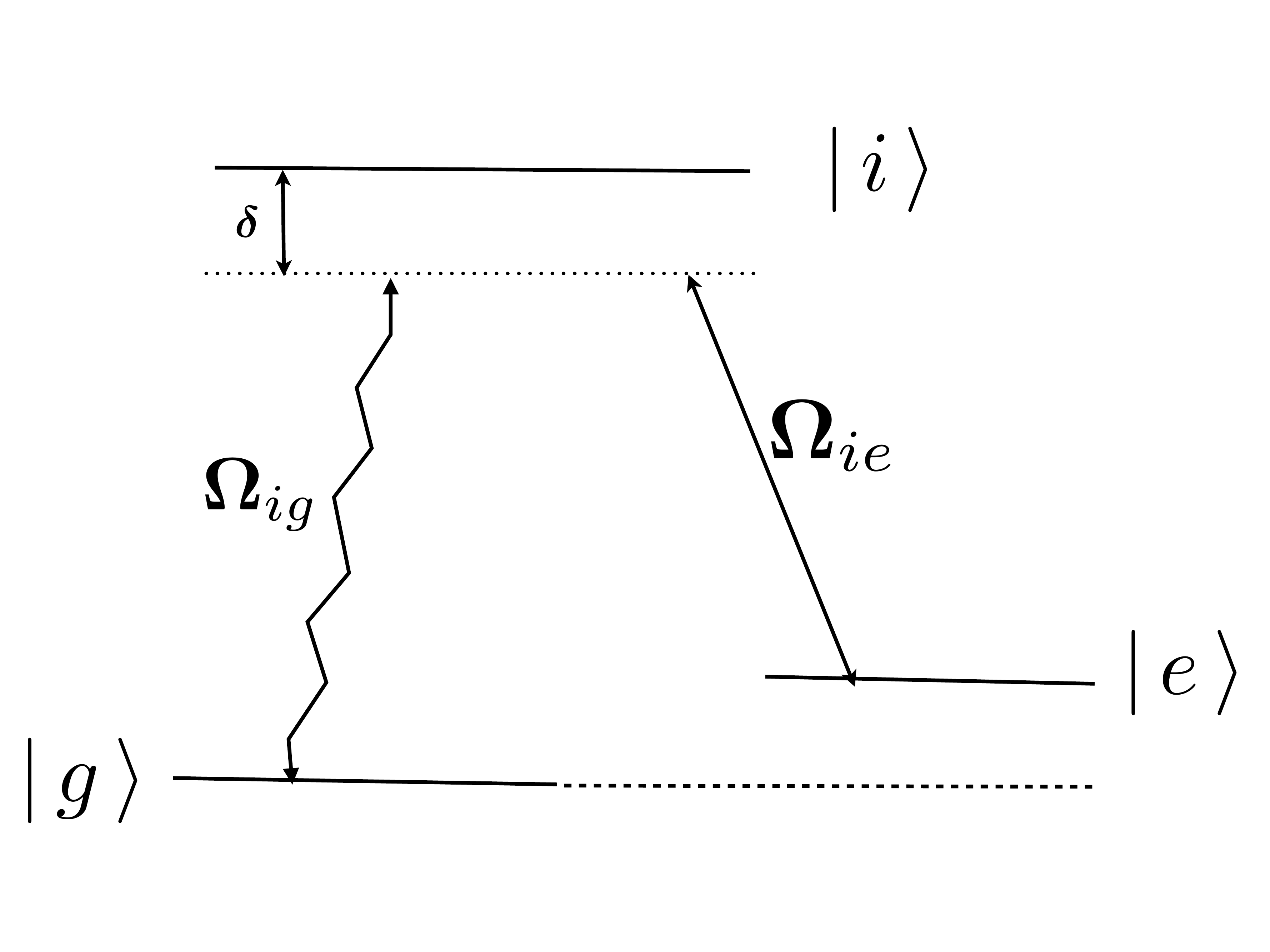}
\caption{Raman transition of a $\Lambda $ atom inside a cavity.}
\end{figure}

Another  example is the case of resonant Raman scattering inside a cavity. Consider a three-level $\Lambda$ atom interacting quasi-resonantly with a mode of frequency $\omega_1$ of the cavity field  and a classical  field of frequency $\omega_2 $, as schematized in Fig~\refc{Raman}.  The two lower levels, $\vert g\rangle$ and $\vert e\rangle,$ are closely spaced in energy and  can make quasi-resonant dipole transitions to an upper level $\vert i\rangle.$ $\omega_{ig}$ and $\omega_{ie}$ are the energy differences between the upper level and the lower levels $\vert g\rangle$ and $\vert e\rangle,$ respectively. 
The Hamiltonian that describes the interaction in the rotating-wave approximation is given by 
\ba\label{Rn}
\widehat H&=&\widehat {H}_0^\prime+\widehat H_{\rm int}^\prime\notag\\
\widehat H_0 ^\prime&=&\omega_{ig }\ketbra{i}{i}+\omega_{eg} \ketbra{e}{e} +\omega_1 \widehat{a}^\dagger \widehat{a}\notag \\
\widehat H_{\rm int}^\prime&=& \Omega_{ig}\ketbra{i}{g}\widehat{a}+ \Omega_{ie}e^{-i\omega_2 t }\ketbra{i}{e} \nonumber\\
&+& \Omega_{ig}\ketbra{g}{i}\widehat{a}^\dagger+ \Omega_{ie}e^{i\omega_2 t }\ketbra{e}{i} ,
\ea
where $\Omega_{ig }$ is the vacuum Rabi frequency associated to the cavity field of frequency $\omega_1$, while $\Omega_{ie }$ is the Rabi frequency associated to the external classical  field of frequency $\omega_2$. 
Assume that the initial cavity field state has a photon distribution with low photon average number.  In Ref.~\cite{Franca01}, it has been shown that if the detuning  $\delta= \omega_{ig}-\omega_1\approx \omega_{ig}-\omega_{eg}-\omega_2$ is such that $|\delta | \gg \Omega _{gi}\gg\Omega _{ei} $,  it is then possible to show that  the Raman transition  $|g,n_0+1\rangle \leftrightarrow |e,n_0 \rangle$ is resonant for a certain $n_0$ depending on the 
detunings of the driving field. Here we rederive the conditions on the frequencies that make the process resonant and the effective hamiltonian for the system.

Assume that  the classical field frequency is tuned to 
\begin{equation}\label{R1}
\omega _{2}=\omega _{1}-\omega_{eg}-(n_{0}+1)\Omega _{gi}^{2}/\Delta  + \Omega _{ei}^{2}/\Delta ,
\end{equation}
with $\Delta $ satisfying the equation
\begin{equation}\label{R2}
\Delta =\omega _{ig}-\omega _{1}+(\Omega _{ei}^{2}+2\Omega
_{gi}^{2})/\Delta +2n_{0}{\Omega _{gi}^{2}}/{\Delta } ,
\end{equation}
where $n_0$ is an integer and $|\Delta | \gg \Omega _{gi}\gg\Omega _{ei} .$  

We first write the Hamiltonian of Eq.~(\ref{Rn}) as $\widehat H=\widehat H_0+\widehat H_{\rm int}$ with
\ba \label{H0corrected}
\widehat H_0&=& \widehat H_0^\prime+\widehat H_{SS}\notag\\
 \widehat H_{\rm int} &=&\widehat H_{\rm int}^\prime -\widehat H_{SS} ,
\ea
where $\widehat H_0^\prime$ and $\widehat H_{\rm int}^\prime$ are given by Eq.~\refc{Rn}. Also, $\widehat H_{\rm SS}$ is given by 
\be
\widehat H_{\rm SS}=(\frac{\Omega
_{gi}^{2}}{\widehat\Delta_g}    \widehat{a} \widehat{a}^\dagger+\frac{\Omega_{ei}^{2}}{\widehat\Delta_e} )\ketbra{i}{i}-\frac{\Omega
_{gi}^{2}}{\widehat\Delta_g }  \widehat{a}^\dagger  \widehat{a}\ketbra{g}{g}- \frac{\Omega_{ei}^{2}}{\widehat\Delta_e }\ketbra{e}{e} \, ,
\ee
where 
\ba
\widehat \Delta_g &=&\omega_1-\omega_{ig}+{\Omega_{ig}^{2}(2\widehat{n}+1)}/{\Delta} \notag \\
\widehat \Delta_e &=&\omega_2-\omega_{ie}+{\Omega_{ie}^{2}\widehat{n}}/{\Delta}\, .
\ea
We then write the time evolution operator in the interaction representation with respect to $\widehat H_0$ and use our unitary perturbative expansion.  Neglecting  terms that vary very rapidly with time, we obtain   
\ba
\lambda \widehat W_1&=&-\widehat H_{SS} t \notag\\
\lambda \widehat W_2&=& \widehat H_{SS}  + \widehat H_{\rm eff} t\, ,
\ea
where   
\be 
\widehat H_{\rm eff} = - \frac{\Omega _{gi} \Omega _{ei}}{\Delta}
(|e\rangle \langle g | \widehat{a} + |g\rangle \langle e|\widehat{a}^{\dagger}).  
\ee
Therefore 
\be
\widehat U_I(t) \approx e^{+i\widehat H_{SS}t}e^{-i(\widehat H_{SS}+\widehat  H_{\rm eff})t}\, .
\ee 

Using the Baker-Hausdorff formula and neglecting the term depending on the commutators of $\wh H_{SS} $ and  $\wh H_{\rm eff}, $ we may write
\begin{equation}
\widehat U_I(t,0) \approx e^{-i\widehat H_{\rm eff}  t} .
\end{equation} 
That is, $\widehat H_{\rm eff}$ can be considered an effective Hamiltonian of the interaction picture associated to the choice of $\widehat{H}_0$ given in the Eq.~\refc{H0corrected}.

 \begin{figure}[ht]
\includegraphics[width=7cm] {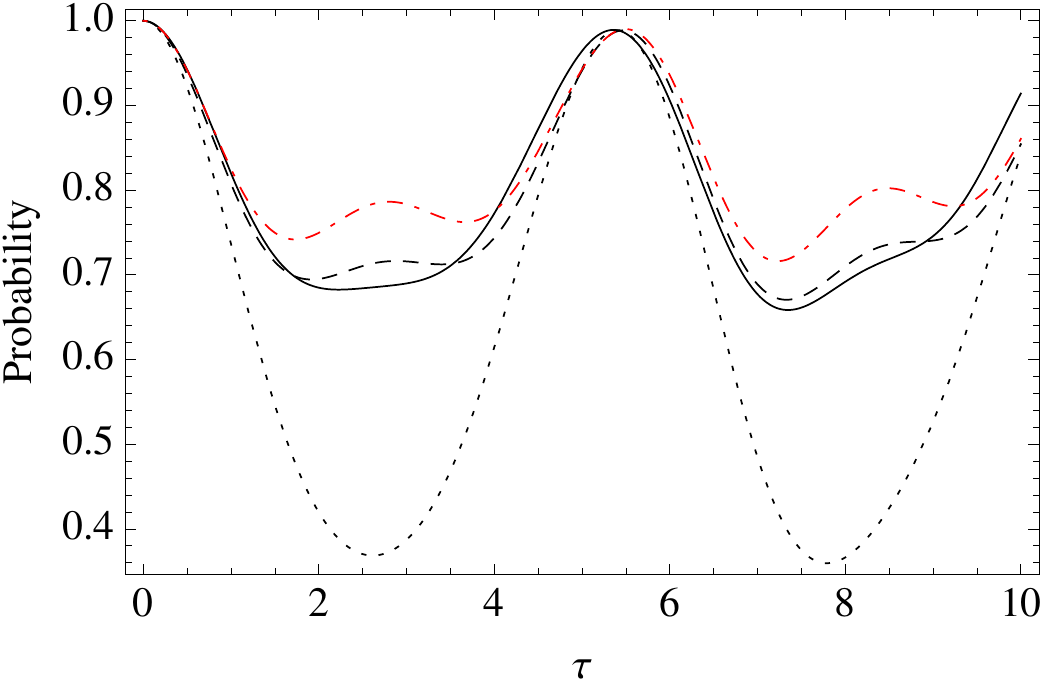}
\caption{{Survival probability of $|g;0\rangle$ vs. $\tau= \omega t.$ Solid line: exact solution; dashed line: $\wh U\approx \wh U_0 \wh U_1; $ dot dashed  line, first Born approx.; dotted line  $\wh U\approx \wh U_0 .$  $\omega_ 0/\omega=0.6; g/\omega=0.5.$}}\label{FigG0W1_Born_H0_x05y06NF20}
\end{figure}

 \begin{figure}[ht]
\includegraphics[width=7cm]{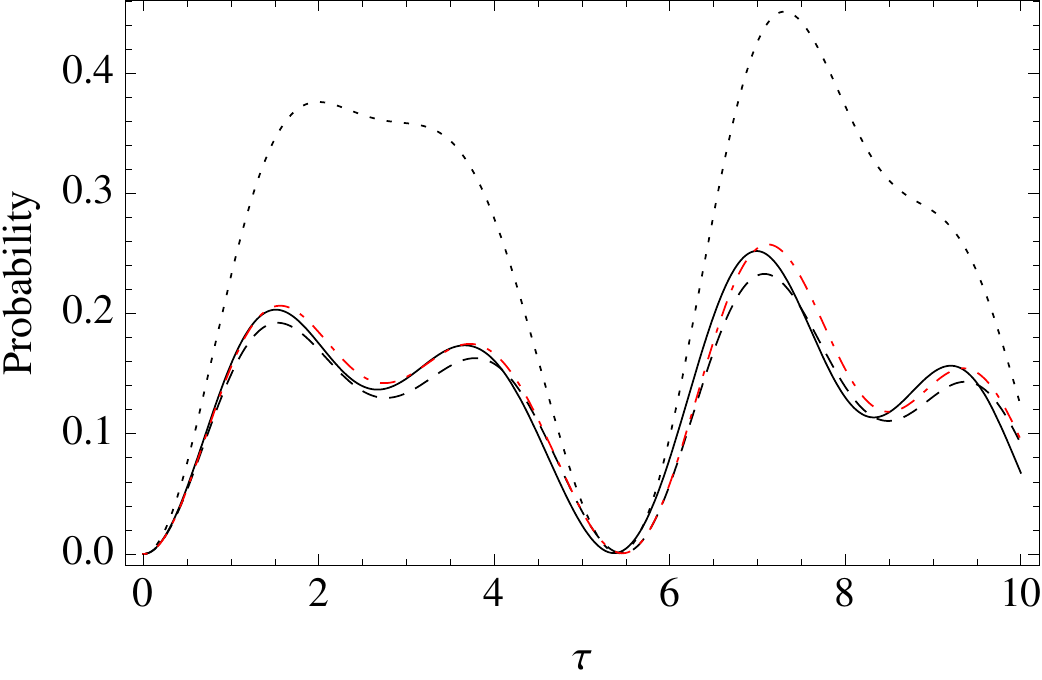}
\caption{{Probability of $|g;0\rangle $ to make a transition to the state $|e;1\rangle$ as a function of $\tau= \omega t.$ Solid line: exact solution; dashed line: $\wh U\approx \wh U_0 \wh U_1;$ dot dashed  line: first Born approximation; dotted line  $\wh U\approx \wh U_0 .$ $\omega_ 0/\omega=0.6; g/\omega=0.5. $}}
\label{FigE1W1_Born_H0_x05y06NF20}
\end{figure}

Consider now the situation  of the Jaynes-Cummings  model in the USC regime between a cavity mode and a qubit, $g / \omega \gtrsim 0.1$. This situation is currently accessible to experiments using superconducting qubits and cavities in circuit quantum electrodynamics~\cite{Niemczyk10,Forn-Diaz10}. In this case, the rotating-wave approximation is no longer valid and  one should consider  the full interaction Hamiltonian
\be
\wh H=\omega \widehat a^\dagger \widehat a+g(\widehat a^\dagger +\widehat a) \widehat \sigma_x +\omega_0 \widehat \sigma_z /2\, .
\ee 
In the case where $\omega_0=0$, it reduces to 
\be \label{Hprime}
\wh H' =\omega \widehat a^\dagger \widehat a+g(\widehat a^\dagger +\widehat a) \widehat \sigma_x\, .  
\ee
The eigenstates of $\wh H'$ are the product of displaced number states~\cite{Oliveira90} and the eigenstates $|\pm \rangle$ of $\widehat \sigma_x$, associated with the eigenvalues $\pm 1$
\be
|\pm n; \pm\rangle = \widehat D(\mp x)|n\rangle\otimes | \pm\rangle ,
\ee
where $x=g/ \omega $, $\widehat D(v)=e^{v \widehat{a}^\dagger-v^* \widehat{a}} $ is the displacement operator,  $|n\rangle , n=0,1,2..$ are Fock states, and  $\widehat\sigma_x |\pm\rangle =\pm |\pm\rangle .$ 
The eigenstates $\widehat  D(\mp x)|n\rangle$  of  $\wh H'$ 
  are degenerated and associated with the eigenvalue $(n\omega-g^2/\omega).$

In basis $\{ |\pm n; \pm\rangle, n=0,1,...\}$,  $\omega_0\widehat \sigma_z/2$ is written as
\be  \label{NRW}
 \omega_0\widehat \sigma_z/2= \omega_0/2 \sum_{n,m} \langle n | \widehat D(2x)| m \rangle \, \,  |+n;+\rangle\langle-m;-| 
  +{\rm   H.c. }
 \ee
 
In Ref.~\cite{Irish05}, it has been proposed an approximation which keeps only the terms with $n=m$ in the right hand side of Eq.~\refc{NRW}, that is,  

 \be  \label{H0GRW}
\wh H_0=\wh H'+
 \omega_0/2 \sum_{n} \langle n | \widehat D(2x)| n \rangle \, \,  |+n;+\rangle\langle-n;-| 
  +{\rm   H.c. }, 
 \ee
 where $\langle n | \widehat D(2x)| n \rangle= e^{-2 x^2} { \cal L}_n(4x^2).$
The eigenstates of $\wh H_0$ can be easily written as
\be 
\frac{1}{\sqrt{2}}(|+ n;+\rangle\pm | -n;-\rangle\, ,
 \ee
and are associated to the eigenvalues $n\omega-g^2/\omega  \pm (\omega_0/2) 
  \langle n | \widehat D(2x)| n \rangle. $

Using the approximate Hamiltonian $\wh H_0$  as our  zeroth-order approximation,  we have found that our method is  well suited for describing transition probabilities  for a very large range of $\omega_0/\omega$ and $g/\omega \gtrsim 0.1$, including both, the USC and DSC regimes of the JC model~\cite{Casanova10}. Let us write $\widehat H = \wh H_0+\wh H_{\rm int} $, where

 \be \label{HIGRW}
 \wh H_{\rm int}=  \omega_0/2 \sum_{n\ne m} \langle n | \widehat D(2x)| m \rangle \, \,  |+n;+\rangle\langle-m;-| 
  +{\rm   H.c. }
 \ee
 
 From Eqs. \refc{H0GRW} and \refc{HIGRW} we can easily calculate 
 $\lambda \wh H_1(t) = e^{i\wh H_0 t} \widehat{H}_{\rm int} e^{-i\wh H_0 t}$ and the operators  $\wh W_1$ and $\wh W_2 $ using the expressions given in Eqs.~\refc{w1} and \refc{w2}.

 \begin{figure}[ht]
\includegraphics[width=7cm] {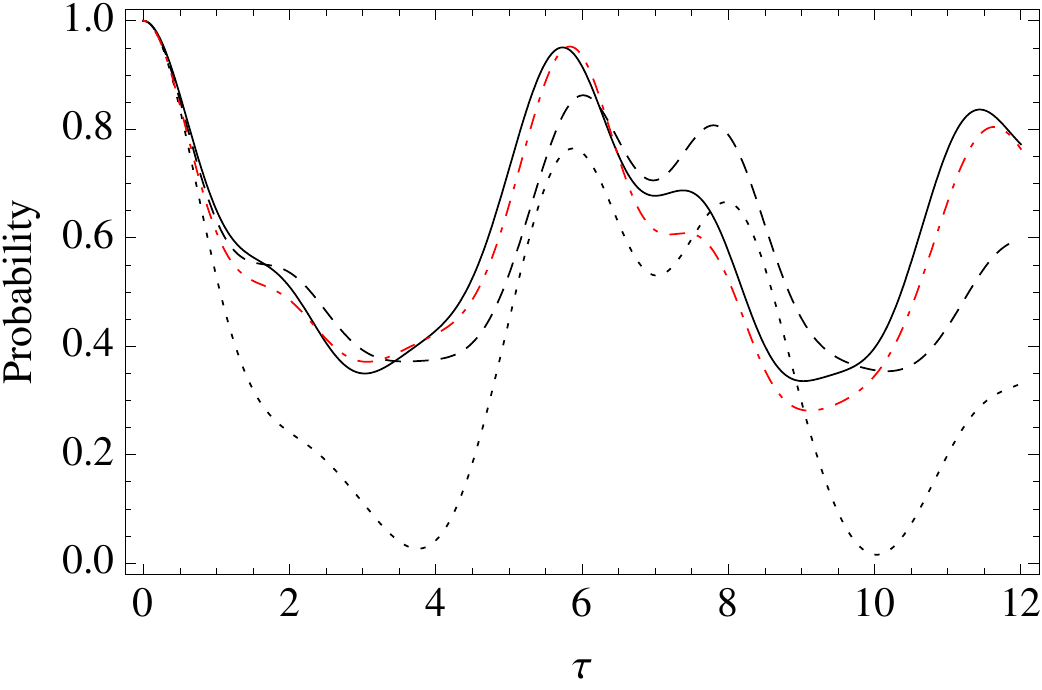}
\caption{{Survival probability of $|g;0\rangle$ as a function of $\tau= \omega t.$ Solid line: exact solution; dashed line: $\wh U \approx e^{-i\wh H_0 t} e^{-i\lambda \wh W_1(t)}; $ dot dashed  line: $\wh U\approx e^{-i\wh H_0 t} e^{-i\lambda \wh W_1(t)}e^{-i\lambda \wh  W_2(t)};  $ 
dotted line  $\wh U\approx  e^{-i\wh H_0 t} .$  $\omega_ 0/\omega=1; g/\omega=0.8. $}}\label{FigG0W2_W1_H0_x08y1NF20}
\end{figure} 

\begin{figure}[ht]
\includegraphics[width=7cm]{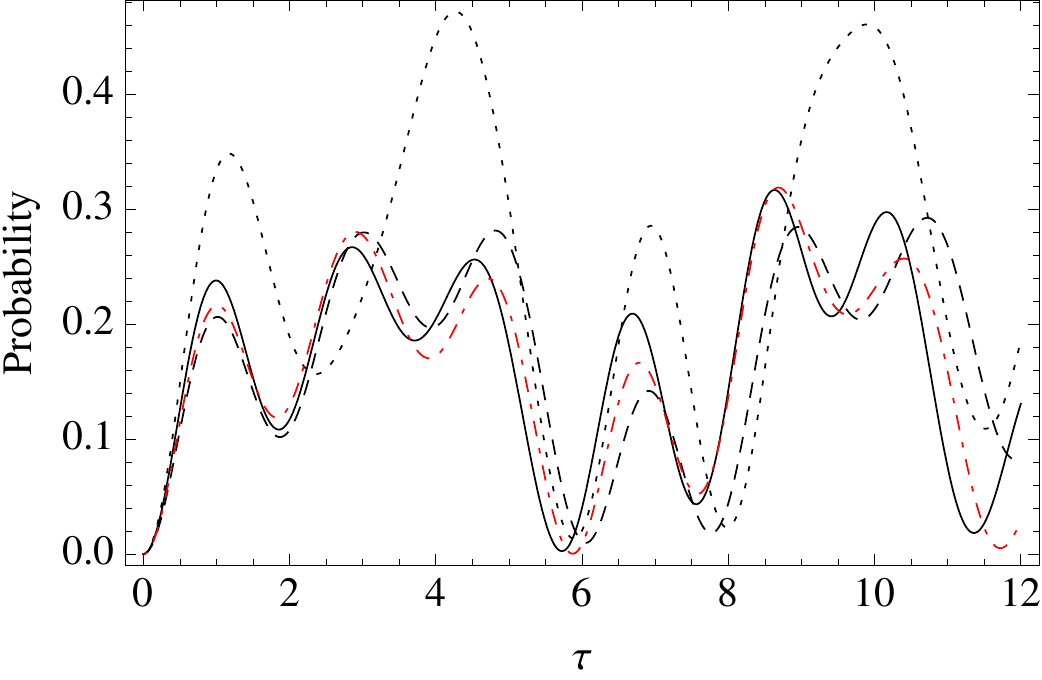}
\caption{{Transition probability from $|g;0\rangle $ to $|e;1\rangle$ vs. $\tau= \omega t.$ Solid line: exact solution; dashed line: $\wh U\approx \wh U_0 \wh U_1; $ dot dashed  line: $\wh U\approx \wh U_0 \wh U_1\wh U_2$; dotted line  $\wh U\approx \wh U_0. $ $\omega_ 0/\omega=1; g/\omega=0.8. $}}\label{FigE1W2_W1_H0_x08y1NF20}
\end{figure}

\begin{figure}[ht]
\includegraphics[width=6.5cm] {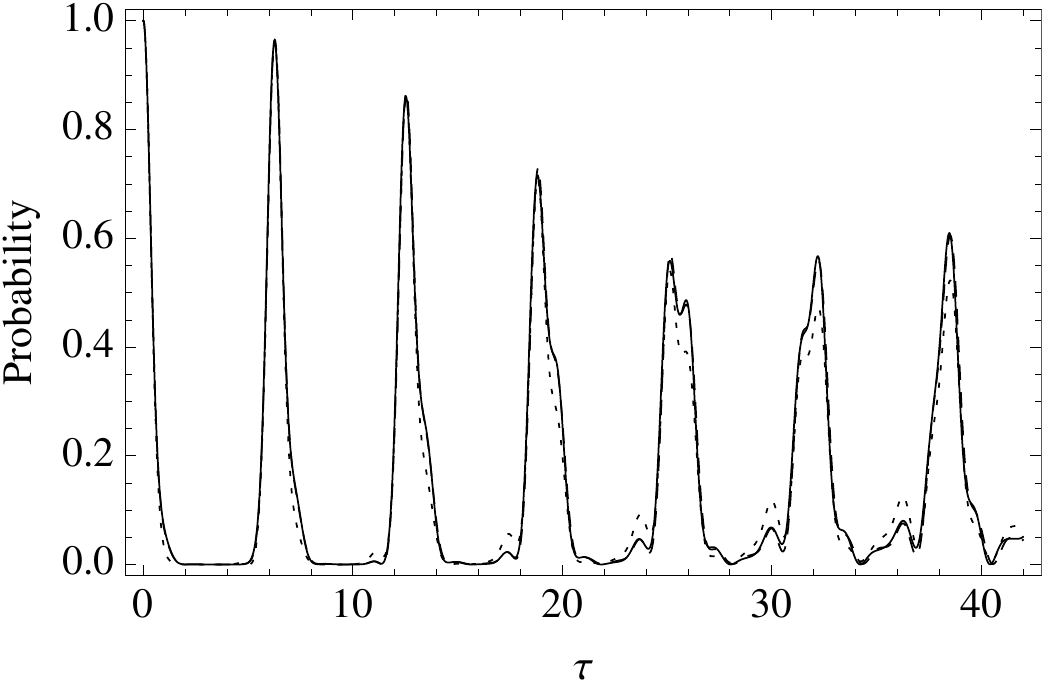}
\caption{{Survival probability of $|g;0\rangle$ as a function of $\tau= \omega t.$ Solid line: exact solution is superposed with $\wh U\approx \wh U_0 \wh U_1; $ dotted line  $\wh U\approx \wh U_0 .$  $\omega_ 0/\omega=0.5; g/\omega=2. $}}\label{Fig6}
\end{figure}

 \begin{figure}[ht]
\includegraphics[width=6.5cm]{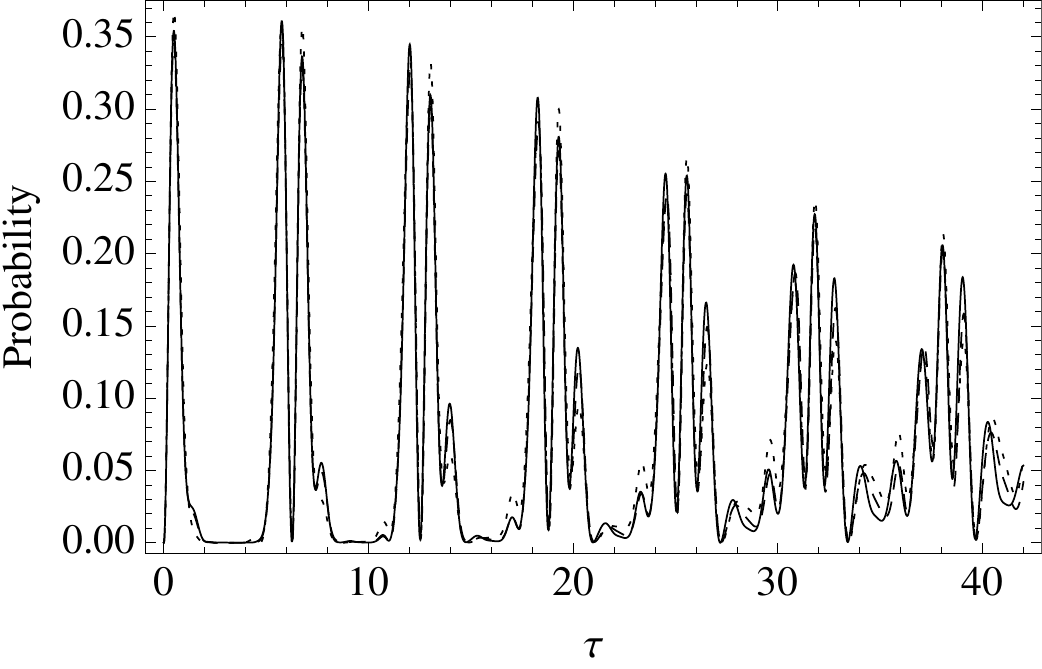}
\caption{{Probability of $|g;0\rangle $ to make a transition to $|e;1\rangle$ as a function of $\tau= \omega t.$ Solid line: exact solution is superposed with $\wh U\approx \wh U_0 \wh U_1; $ dotted line  $\wh U\approx \wh U_0 .$  $\omega_ 0/\omega=0.5; g/\omega=2. $}}
\label{Fig7}
\end{figure}
 
 In Figs.~\ref{FigG0W1_Born_H0_x05y06NF20} and \ref{FigE1W1_Born_H0_x05y06NF20},
 we present curves corresponding to the exact result calculated numerically, the results calculated using three approximations for the time evolution operator: $  e^{-i\wh H_0 t}, $ $ e^{-i\wh H_0 t} e^{-i\lambda \wh W_1(t)}, $ and the Dyson approximation to first order in $\omega_0$. The results show that both, the approximation $\widehat H=\widehat H_0$ and the first Born approximation, do not describe the transition probabilities for the USC regime. On the other hand, our unitary expansion describes quite well the results, even when we take only the first non-trivial contribution. Note that the contribution of $\wh W_2$ is not shown and gives a small correction. 
 
In Figs.~\ref{FigG0W2_W1_H0_x08y1NF20} and \ref{FigE1W2_W1_H0_x08y1NF20}, we also present curves corresponding to the exact result calculated numerically and  those calculated using three approximations for the time evolution operator: $  e^{-i\wh H_0 t}, $ $ e^{-i\wh H_0 t} e^{-i\lambda \wh W_1(t)} $ and  $ e^{-i\wh H_0 t} e^{-i\lambda \wh W_1(t)}e^{-i\lambda \wh  W_2(t)}.  $  The first Born approximation is not shown and represents a small correction to the approximation $\widehat H\approx \widehat H_0.$ We see that, although the terms associated with $\wh W_1$ gives the main contribution, the terms associated with $\widehat{W}_2$ does improve the approximation. These considerations are also applicable to the DSC regime~\cite{Casanova10}, as can be seen in Figs.~6 and 7.

\section{Conclusions}
We introduced a perturbative unitary expansion for the time evolution operator as a product of exponentials of antihermitian operators. Consequently, this expansion can be truncated at any order of approximation while keeping unitarity.  We have presented three examples: a harmonic oscillator with a time-dependent force,  the Raman transition inside a resonant cavity, and  the James-Cummings model in the USC and DSC regimes.

\section*{Acknowledgments}
J.C. acknowledges funding from Basque Government BFI08.211,  E.S. from Basque Government Grant IT472-10, Spanish MICINN FIS2009-12773-C02-01, and SOLID European project, and  N.Z. from Brazilian agencies CNPq and FAPERJ. N.Z. would like to thank Prof. E. Solano and the Universidad del Pa\'{\i}s Vasco for hospitality.


\begin{thebibliography}{99}

\bibitem{Cohen91}
 C. Cohen-Tannoudji, B. Diu, and F. Lalo\"e, {\it Quantum Mechanics}, Vol. 1 (Wiley, New York, 1991).
 
\bibitem{Dyson49}
F. J. Dyson, Phys. Rev. {\bf 75}, 486 (1949), F. J. Dyson, Phys.Rev {\bf 75}, 1736 (1949).

\bibitem{Magnus54}
W. Magnus, Commun. Pure Appl. Math. {\bf 7}, 649 (1954).

\bibitem{Fer58}
F.~Fer, Bull. Classe Sci. Acad Roy. Bel. {\bf 44}, 818 (1958)

\bibitem{Aniello05}
P. Aniello, J.Opt. B: Quantum Semiclass. Opt. {\bf 7}, S507 (2005).

\bibitem{Blanes98}
S. Blanes, F, Casas, J.~A.~Oteo and J.~Ros , J. Phys. A {\bf 31}, 259 (1998).

\bibitem{see}
N. Wiebe, D. W. Berry, P. Hoyer, and B. C. Sanders, J. Phys. A: Math. Theor. {\bf 43}, 065203 (2010).

\bibitem{Pechugas66}
P. Pechugas and J. C. Light, J. Chem. Phys. {\bf 44}, 3897 (1966).

\bibitem{Franca01}
M. F. Santos, E. Solano, and R. L. de Matos Filho, Phys Rev Lett. {\bf 87}, 093601 (2001).

\bibitem{Niemczyk10}
T. Niemczyk, F. Deppe, H. Huebl, E. P. Menzel, F. Hocke, M. J. Schwarz, J. J. Garc\'{\i}a-Ripoll, D. Zueco, T. H\"ummer, E. Solano, A. Marx, and R. Gross, Nature Phys. {\bf 6}, 772 (2010).

\bibitem{Forn-Diaz10}
P. Forn-D\'iaz, J. Lisenfeld, D. Marcos, J. J. Garc\'{\i}a-Ripoll, E. Solano, C. J. P. M. Harmans, and J. E. Mooij, Phys Rev Lett. {\bf 105}, 237001 (2010).

\bibitem{Oliveira90}
S. M. Roy and V. Singh, Phys. Rev. {\bf 25}, 3413 (1982); F. A. M. deOliveira, M. S. Kim, P. L. Knight, and V. Bu\v zek, Phys. Rev. A, {\bf 41}, 2645 (1990).

\bibitem{Irish05}
E. K. Irish, J. Gea-Banacloche, I. Martin, and K. C. Schwab, Phys. Rev. B {\bf 72}, 195410 (2005).

\bibitem{Casanova10}
J.~Casanova, G. Romero, I. Lizuain, J. J. Garc\'{\i}a-Ripoll, and E. Solano,  Phys Rev Lett. {\bf 105}, 263603 (2010).


\end{thebibliography}
\end{document}